\documentclass{aa}  

\usepackage{graphicx}
\usepackage{txfonts}
\usepackage{xcolor}
\newcommand{\msun}{\rm\, M_\odot}

\begin{document}

   \title{Size matters: are we witnessing super-Eddington accretion in high-redshift black holes from JWST?}

   %\subtitle{I. Overviewing the $\kappa$-mechanism}

   \author{Alessandro Lupi
          \inst{1,}\
          \inst{2}\fnmsep\thanks{alessandro.lupi@uninsubria.it}
          \and
          Alessandro Trinca
          \inst{1,}\
          \and
          Marta Volonteri
          \inst{3}\
          \and
          Massimo Dotti
          \inst{4,}\
          \inst{2}
          \and
          Chiara Mazzucchelli
          \inst{5}
          }

   \institute{Dipartimento di Scienza e Alta Tecnologia, Universit\`a degli Studi dell'Insubria, via Valleggio 11, I-22100, Como, Italy
        \and
            INFN, Sezione di Milano-Bicocca, Piazza della Scienza 3, I-20126 Milano, Italy
         \and 
            Institut d’Astrophysique de Paris, UMR 7095, 
                CNRS and Sorbonne Universit\'{e}, 98 bis boulevard Arago, 75014 Paris, France
        \and
        Dipartimento di Fisica ``G. Occhialini'', Universit\`a degli Studi di Milano-Bicocca, Piazza della Scienza 3, I-20126 Milano, Italy
        \and
        Instituto de Estudios Astrof\'isicos, Facultad de Ingenier\'ia y Ciencias, Universidad Diego Portales, Avenida Ejercito Libertador 441, Santiago, Chile
             }

   \date{Draft \today }

% \abstract{}{}{}{}{} 
% 5 {} token are mandatory
 
  \abstract{
  Observations by the James Webb Space Telescope of the Universe at $z\gtrsim 4$ have shown that massive black holes (MBHs) appear extremely overmassive compared to the local correlation for active galactic nuclei. In some cases, these objects might even reach half the stellar mass inferred for the galaxy. Understanding how such objects formed and grew to this masses has then become a big challenge for theoretical models, with different ideas ranging from heavy seed to super-Eddington accretion phases. Here, we take a different approach, and try to infer how accurate these MBH mass estimates are and whether we really need to revise our physical models. By considering how the emerging spectrum (both the continuum and the broad lines) of an accreting MBH changes close to and above the Eddington limit, we infer a much larger uncertainty in the MBH mass estimates relative to that of local counterparts, up to an order of magnitude, and a potential preference for lower masses and higher accretion rates, which i) move them closer to the local correlations, and ii) might indicate that we are witnessing for the first time a widespread phase of very rapid accretion.  
  }
  

   \maketitle
%
%________________________________________________________________

\section{Introduction}
Massive black holes (MBHs) are ubiquitously found to inhabit the centre of massive galaxies up to redshift $z\gtrsim 6$ \citep[e.g.][]{fan06,mortlock11,banados18,fan23,maiolino23}, with masses in the range $\sim 10^5-10^{10}\msun$. Observationally, they are commonly identified via gas accretion through the conversion of gravitational energy into radiation, which makes them shine as Active Galactic Nuclei (AGN). They sometimes also produce powerful collimated jets.

MBHs are expected to gain most of their mass via radiatively efficient accretion \citep{soltan82,marconi2004}, hence they should have formed from lower-mass black hole `seeds' (see, e.g., \citealt{Inayoshi20} and \citealt{Volonteri21} for a review).

With the advent of the James Webb Space Telescope (JWST), we have pushed the observational limit deeply in the dark ages, finding galaxies up to $z\sim 14$ \citep{carniani24}. Some of these galaxies also hosted MBHs, which appear to challenge most MBH formation mechanisms, unless one assumes an initially heavy seed ($10^4-10^5\msun$) and a continuous growth at the Eddington limit. Theoretical models are further challenged by the large abundance of these (candidate) objects \citep[e.g.][]{harikane23,maiolino23,greene24}, which implies a formation efficiency of massive seeds way larger than what found in theoretical models. Several studies have shown that this issue can be alleviated if one considers the plausibility of accretion above the Eddington limit \citep[e.g.,][]{lupi16,pezzulli16,regan19,massonneau23,lupi24,shi24}, which can compensate for the stunted growth in low-mass galaxies \citep[see, e.g.][]{anglesalcazar17FIRE}. Apart from the mass of these MBHs, another important difference with the local population is that these MBHs seem to be extremely massive compared to their galaxy hosts, lying well above the local correlations \citep{farina22,maiolino23,yue24,stone24} in some cases even weighing more than half the total stellar mass \citep{Juodzbalis24}. Note, however, that an important role in the comparison is also played by the galaxy mass employed, either the stellar mass, as in the recent JWST results, or the dynamical mass, as in the case of ALMA observations \citep{decarli18,izumi21,farina22}. Among the theoretical efforts to explain these systems, the most promising solutions consider i) a somewhat extremely efficient heavy seed formation at high redshifts and an efficient suppression of star formation by the accreting MBH, ii) a strong observational bias \citep{Li2024}, or iii) a population of primordial MBHs \citep{ziparo22,dolgov24}. Despite these efforts, a clear consensus is still missing to date, partially because of the large uncertainties in the stellar mass and (potentially) in the MBH mass.
   This second possibilty is rarely considered. In fact, the MBH mass at these redshifts is commonly inferred through the single-epoch method, employing the virial theorem combined with the correlations between the broad H$\alpha$, H$\beta$, or MgII line widths and luminosities, and the emission properties of the continuum emitted by the innermost regions of the accretion disc. These correlations have been calibrated in the local Universe  \citep[$z\lesssim 0.3$][]{vestergaard09,bentz13,reines13,rv15}, and then extrapolated at high redshift.

   Recently, \citet{king24} pointed out that close-to-Eddington or super-Eddington accreting MBHs would have $(i)$ the emission from the accretion disc beamed by multiple scatterings within the funnel created by a central thickening of the disc itself and $(ii)$ unvirialized BLR whose dynamics is mostly dominated by outflows. Under such conditions, \citet{king24} demonstrated that the MBH estimates inferred would be artificially biased towards high values, and argued that such an effect might be particularly relevant for high-redshift AGN. Another potential source of bias could instead result from an inaccurate estimate of the broad line region (BLR) size, as suggested by recent reverberation mapping campaigns - including the SEAMBH \citep{du2014} and the SDSS-RM \citep{grier2017} - of multiple highly accreting MBHs, \cite{MartinezAlmada2019}.
   
   In particular, these campaigns demonstrated that the time lag of the H$\beta$ line, which is directly associated to the size of the BLR, depends on the accretion rate of the MBH, and shortens for accretion rates above $f_{\rm Edd}\sim 0.3$ \citep[][W14 hereafter]{wang14c}. The proposed interpretation for such effect is radiation pressure which, for accretion rates close and above the Eddington limit, thickens the accretion disc. Such a thicker disc is better described by the slim-disc solution \citep{abramowicz88} rather than by a more standard radiatively efficient \citet[][SS hereafter]{shakura73} disc, and results in a lower flux of ionizing photons reaching the BLR clouds compared to a radiatively efficient AGN with identical optical spectrum. In these conditions, the BLR splits in unshadowed and shadowed regions, the latter receiving less photons and shrinking in size, which result in a net shorter lag. 
   
   Motivated by these results, in this work we explore the effect of a varying BLR size, based on the aforementioned results and on a fully physical approach, on the inferred MBH masses in the most challenging high-redshift sources observed by JWST to date. In particular, we account for the possibility that the observed luminosities might be the result of a lower mass, highly accreting MBH, with the aim of assessing potential biases in the MBH mass estimates.
   
   The manuscript is organised as follows. In Section~\ref{sec:methods} we describe our procedure to estimate the MBH mass, in Section~\ref{sec:results} we present our results, and in Section~\ref{sec:discussion} we discuss potential caveats in the analysis and draw our conclusions.

%__________________________________________________________________

\section{Methods}
\label{sec:methods}
In order to test how relevant the evolution of the BLR size with the Eddington ratio is in high-redshift systems, we build a theoretical model of the accretion disc and the BLR emissions based on the electromagnetic spectrum of a slim disc, as defined by the \textsc{agnslim} model in \textsc{xspec} \citep{kubota19}. Of the many parameters available in the model, in our work we only considered the impact of the three main ones: the MBH mass $M_{\rm BH}$, the Eddington ratio $L_{\rm thin}/L_{\rm Edd}\equiv \eta_{\rm thin}\dot{M}_{\rm BH}c^2/L_{\rm Edd}$, with $L_{\rm thin}$ and $\eta_{\rm thin}$ being the bolometric luminosity and the radiative efficiency of a SS disc, and the MBH spin $a_{\rm BH}$, leaving the others to their default value. We sample 6250 different combinations, with 25 logarithmically-spaced MBH masses between $10^5$ and $10^{10}\msun$, 25 logarithmically-spaced Eddington ratios in the range $0.01 - 10^3$, and 10 linearly-spaced values of the MBH spin between $0$ and $0.998$. The spectrum covers the energy range $0.1$~eV -- $100$~keV, corresponding to a wavelength range $0.12$\AA\ -- $12.4
\,\mu$m, in 1000 logarithmically-spaced bins. 

After the spectra have been generated, for each combination we tabulate the luminosity at 5100\AA\ ($\lambda L_\lambda$), and the ionising luminosity $L_{\rm ion}$ above $E>0.1$~keV \citep[soft-X;][]{kwan79}, the latter needed to determine the broad-line emission from the disc properties \citep[see, e.g.][]{osterbrock06}. For consistency with \citet{kubota19}, we normalize the spectrum bolometric luminosity to the value estimated from the numerical integration of the slim disc solution by \citet{sadowski11}. This normalisation allows gives us the effective radiative efficiency $\eta$ for each combination of the three model parameters, that we use in the rest of the paper to determine $L/L_{\rm Edd}=\eta/\eta_{\rm thin} L_{\rm thin}/L_{\rm Edd}$.
With the table so created, we then built a theoretical model for the BLR emission to be compared with observations. In particular, the observed quantities we considered are: the broad-line width (either H$\alpha$ or H$\beta$) and the luminosity (either the H$\alpha$ luminosity or the luminosity at 5100\AA), according to the values reported in the corresponding observational works \citep{harikane23,maiolino23,ubler23,greene24}, both with their associated uncertainties $\sigma$. 

Our model is defined as follows.
\begin{itemize}
    \item Given a specific combination of $M_{\rm BH}$, $L_{\rm thin}/L_{\rm Edd}$ and $a_{\rm BH}$, we extract $L_{\rm 5100\mathring{A}}$ and $L_{\rm ion}$ via tri-linear interpolation on our table. 
    
    \item For simplicity, we do not make any specific assumption on the cloud properties in the BLR, and generically assume that they are homogeneously distributed around the central MBH \citep{wang14c}.\footnote{This is a very simplistic assumption, as both the cloud angular distribution and their maximum distance from the source are completely unconstrained. Previous studies hinted at a common disc-like geometry for the BLR \citep[e.g.][]{Wills86, Collin87, Runnoe13}. We stress that a flatter BLR would enhance the self-shadowing effect.} Following W14, we assume that self-shadowing is negligible within the funnel, which is defined by an aperture
    \begin{equation}
        \theta_{\rm fun}\approx\left\{
        \begin{array}{lc}
            90^\circ & f_{\rm Edd}<8 \\
            118^\circ-33^\circ\log{f_{\rm Edd}} & 8\geq f_{\rm Edd}<100\\
            76^\circ-12^\circ\log{f_{\rm Edd}} & f_{\rm Edd}\geq 100
        \end{array}\right.,
    \end{equation} where $f_{\rm Edd} \equiv  \dot{M}_{\rm BH}c^2/L_{\rm Edd} = \eta_{\rm thin}^{-1}L_{\rm thin}/L_{\rm Edd}$, and that the ionising radiation emitted within this solid angle directly impinges on the BLR clouds. Assuming an intrinsic spectrum with angular distribution $dF/d\theta \propto \cos\theta$, we then determine the broad-line emission from clouds within the funnel solid angle assuming the local correlation \citep{greene05}
    % \begin{equation}
    %     L_{\rm H\beta} = L_{\rm H\beta,ref}(N_{\rm ion}/N_{\rm ion,ref}),
    % \end{equation}
    % where $N_{\rm ion,ref}$ is the number of ionising photons for a standard SS disc, that we obtain from our table assuming $L/L_{\rm Edd}=0.01$, and $L_{\rm H\beta,ref}$ is defined as 
    \begin{equation}
        \frac{L_{\rm H\beta,fun}}{10^{42}\rm\, erg\, s^{-1}} = (1.425\pm 0.007)\left(\frac{x_{\rm fun}L_{\rm 5100\mathring{A}}}{10^{44}\rm\, erg\, s^{-1}}\right)^{1.133\pm 0.005},
        \label{eq:LHb}
    \end{equation}
    where $x_{\rm fun}$ is the fraction of the total ionising flux within the funnel.
    Outside the funnel, instead, we model self-shadowing through Eq.~(19) in W14
    \begin{equation}
        \frac{L_{\rm H\beta,s-s}}{L_{\rm H\beta,fun}}\approx 0.28\frac{\xi_{\rm s-s}}{\xi_{\rm fun}}\frac{\cos\theta_{\rm fun}}{1-\cos\theta_{\rm fun}}\left(\frac{f_{\rm Edd}}{50}\right)^{-0.6},
    \end{equation}
    where $\xi_{\rm s-s}$ and $\xi_{\rm fun}$ are the anisotropic factors for H$\beta$ emission from the BLR clouds in the self-shadowed region and within the funnel respectively. As these values are completely unconstrained, but for pole-on observers (where they are both equal to unity). Here, we assume for simplicity that they are always of the same order and remove them from the equation.
    The total H$\beta$ luminosity is finally estimated as $L_{\rm H\beta}=L_{\rm H\beta,fun}+L_{\rm H\beta,s-s}$.
    
    In order to determine the H$\alpha$ luminosity, we  assume the standard scaling from \citet{greene05}
    \begin{equation}
        \frac{L_{\rm H\alpha}}{10^{42}\rm\, erg\, s^{-1}}  = (5.25\pm 0.02)\left(\frac{L_{\rm 5100\mathring{A}, proxy}}{10^{44}\rm\, erg\, s^{-1}}\right)^{1.157\pm 0.005},
        \label{eq:LHa}
    \end{equation}
    where  
    \begin{equation}
        \frac{\tilde{L}_{\rm 5100\mathring{A},proxy}}{10^{44}\rm\, erg\, s^{-1}}= \left(\frac{L_{\rm H\beta}}{(1.425\pm 0.007)\times 10^{42}\rm\, erg\, s^{-1}}\right)^{1/(1.133\pm 0.005)}.
    \end{equation}
    We note that, when the broad-line flux or luminosity are not reported, as in the case of \citet{yue24}, we directly compare $L_{\rm 5100\mathring{A}}$ from our model with the observed data.
     
     \item The last piece of information we need for the model is the full-width-half-maximum ($FWHM$) of the broad lines, which we determine by assuming virial equilibrium in the BLR, which gives
     \begin{equation}
         FWHM_{\rm H\beta}= \sqrt{\frac{R_{\rm BLR}}{f_{\rm virial}GM_{\rm BH}}}
         \label{eq:fwhmHb}
     \end{equation}
     for the H$\beta$ line, where $f_{\rm virial}$ is a parameter taking into account the unknown inclination, geometry, and kinematics of the BLR. In this work, we consider as our `fiducial' case $f_{\rm virial}= 1.075$ \citep{rv15}, but also explore a case in which $f_{\rm virial}\propto (FWHM_{\rm line, obs})^{-k}$ \citep[][MR18 hereafter]{mejiarestrepo18}, with $k=1$ (H$\alpha$) or $k=1.17$ (H$\beta$).
     In order to estimate $R_{\rm BLR}$, we employ the relations derived by \citet{MartinezAlmada2019}
     \begin{equation}
         \log{\frac{R_{\rm BLR}}{R_{\rm BLR,Ref}}} = \alpha \log{f_{\rm Edd}} + \beta,
     \end{equation}
     which takes into account the self-shadowing of the BLR. For the fiducial model, we set $\alpha=-0.143 $, $\beta=-0.136$, and assume $R_{\rm BLR,Ref}$ as the reference H$\beta$ BLR size estimate by \citet{bentz13}
     \begin{equation}
         \log{\frac{R_{\rm BLR,Ref}}{\rm 1 lt-day}} = 1.527\pm 0.31 + 0.533^{+0.035}_{-0.033}\log{\frac{L_{\rm 5100\mathring{A}}}{10^{44}\rm\, erg\, s^{-1}}}.
     \end{equation}
     In the MR18 case, we employ instead $\alpha=-0.283$, $\beta=-0.228$, and $f_{\rm Edd}= f_{\rm virial}^{-2}\eta_{\rm thin}^{-1}L_{\rm thin}/L_{\rm Edd}$.
     For sources where the MBH mass is estimated from the $H\alpha$, we finally convert $FWHM_{\rm H\beta}$ to the $FWHM_{\rm H\alpha}$  through the \citet{bentz13} relation
     \begin{equation}
         FWHM_{\rm H\beta}= (1.07\pm 0.07)\times 10^{3}\left(\frac{FWHM_{\rm H\alpha}}{10^3\rm\, km\, s^{-1}}\right)\rm\, km\, s^{-1}.
     \end{equation}
     
\end{itemize}

In order to compare our model predictions with observations, we employ a Markov-Chain Monte Carlo (MCMC) algorithm as implemented in the \textsc{emcee} package \citep{foremanmackey13}. We consider here as our observational sample the sources identified by \citet{harikane23,maiolino23,yue24,ubler23}, and \citet{greene24}, in the redshift range $4\lesssim z\lesssim 7$. The likelihood $\mathcal{L}$ for the MCMC is defined through 
\begin{equation}
    \ln \mathcal{L} = -\frac{1}{2} \sum_i \left[ \frac{(Y_i-\bar{Y}_i)^2}{s_i^2} + \ln (2\pi s_i^2)\right],
\end{equation}
where $Y_i$ is the observed broad-line $FWHM$ and the luminosity, $\bar{Y}_i$ is the value predicted by our model, and $s_i$ is the uncertainty in the observed data (assumed Gaussian). The parameters of our model that we aim at constraining are %Such a likelihood function is simply a Gaussian with the uncertainty underestimated by a fraction $f$, which we optimize together with the other parameters of the model 
$M_{\rm BH}$, $L_{\rm thin}/L_{\rm Edd}$, and $a_{\rm BH}$. As priors, we assume a log-flat distribution for $M_{\rm BH}$ and $L_{\rm thin}/L_{\rm Edd}$ over the intervals $[5,10]$ and $[-3,3]$ respectively, and a uniform distribution for $a_{\rm BH}$ between 0 and 0.998. %Finally, we set a log-flat distribution for $f$ over the range $-10 - 1$. 
We ran the MCMC for 10000 steps employing 32 walkers.\footnote{The number of steps chosen corresponds to about 100 autocorrelation time-scales, which is sufficient to guarantee a robust optimisation.} In order to incorporate the uncertainties in the correlations used by our model, every time we employ one of the relations above, we sample the slope and normalisation from a Gaussian distribution centred on the best-fit value and with $\sigma$ defined by the uncertainty of the fit.\footnote{When the uncertainties are asymmetric, we approximate the distribution as a Gaussian distribution with $\sigma_{\rm eff}$ the average between the two uncertainties.} This choice ensures a proper coverage of the parameter space, even with a very limited dataset given by only two values. In the case of a $FWHM$-dependent virial factor, we randomly sample the virial factor for each source before starting the MCMC from a Gaussian distribution centred on the observed broad-line $FWHM$ with the observed uncertainty, and keep it constant throughout the optimisation procedure, in line with the correlation found by \citet{mejiarestrepo18}. 

\section{Results}
\label{sec:results}
\subsection{Model validation}

\begin{figure}
    \centering
    \includegraphics[width=\columnwidth]{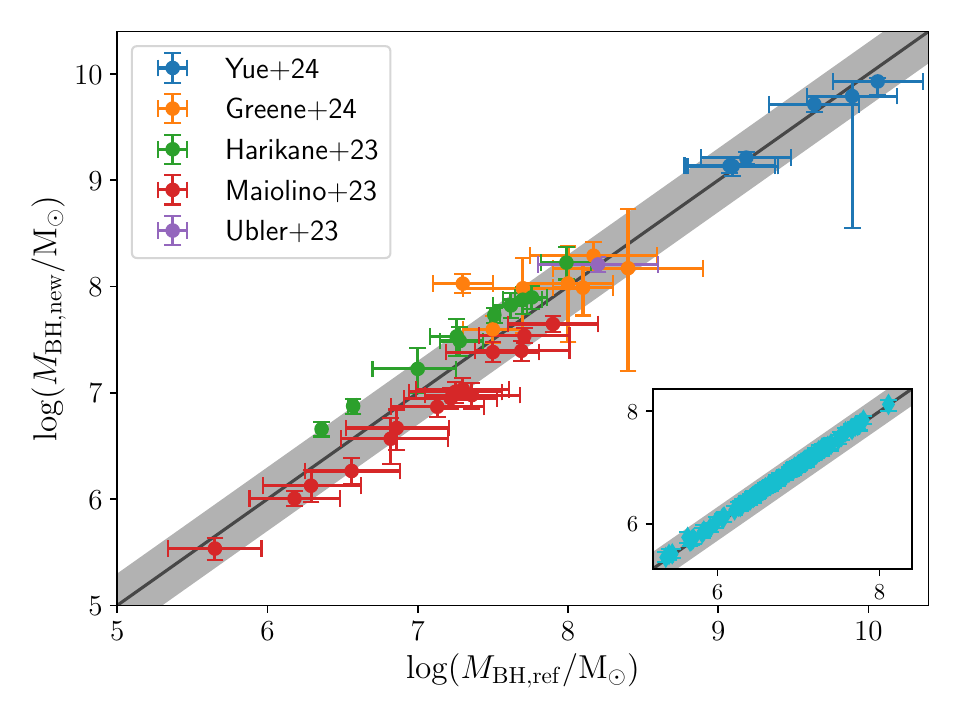}
    \caption{$M_{\rm BH}$ estimates from the MCMC for the validation run against the MBH mass reported in the observational studies considered in this work. The black line corresponds to the 1:1 relation, with the grey shaded area  0.3~dex wide. The dots correspond to the MBHs in \citet[][blue]{yue24}, \citet[][orange]{greene05}, \citet[][green]{harikane23}, \citet[][red]{maiolino23}, and \citet[][purple]{ubler23}. In the inset we show the results obtained for the \citet{rv15} data as cyan crosses.}
    \label{fig:MBH}
\end{figure}

\begin{figure*}
    \centering
    \includegraphics[width= 0.95\textwidth]{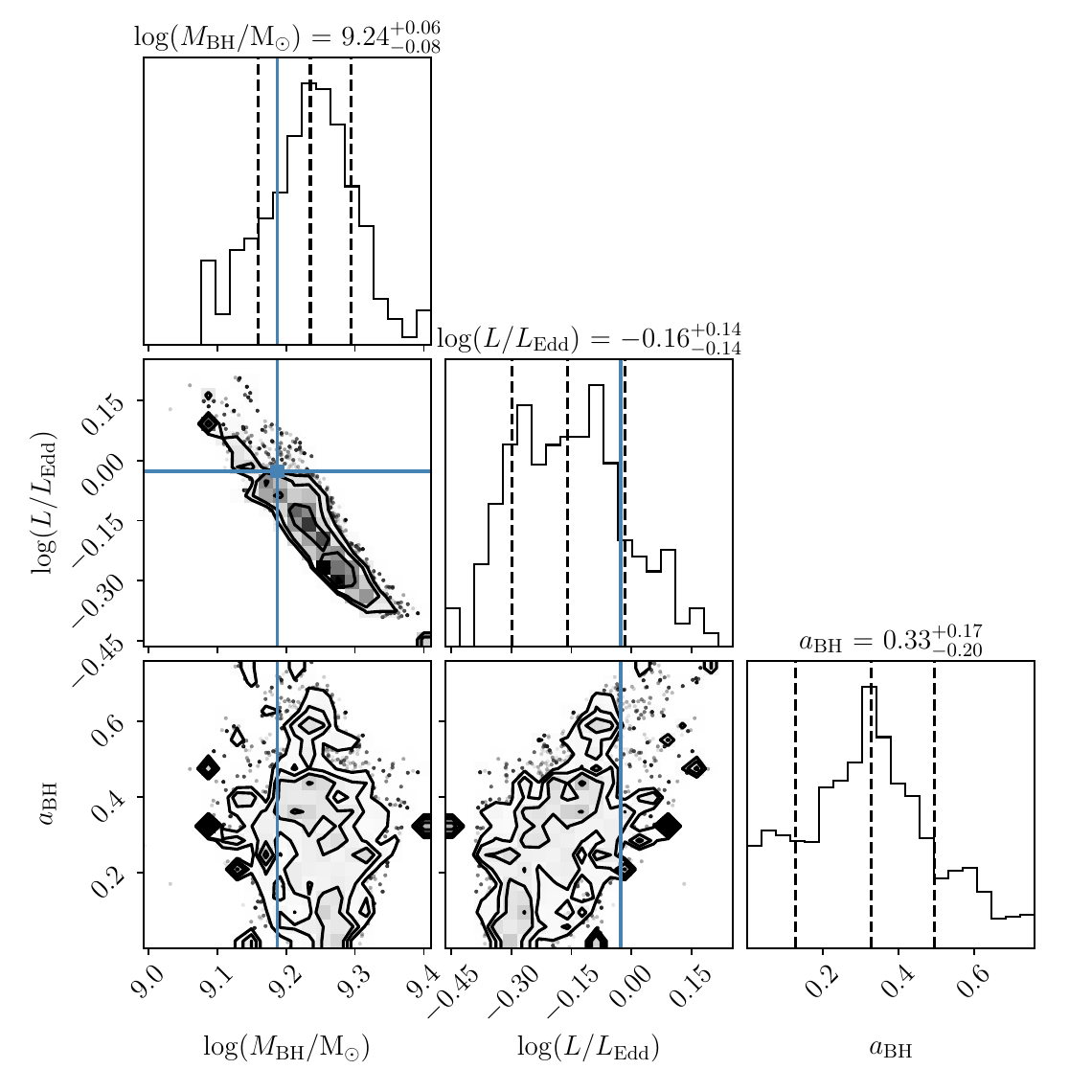}
    \caption{Corner plot resulting from the MCMC validation run on J1030+0524 \citep{yue24} for the three physical parameters of the model $M_{\rm BH}$, $L/L_{\rm Edd}$ (obtained by rescaling $L_{\rm thin}/L_{\rm Edd}$ as described in Section~\ref{sec:methods}), and $a_{\rm BH}$. The blue lines correspond to the values reported in the original work.}
    \label{fig:corner}
\end{figure*}

Before running the MCMC with the fiducial model described in the previous section, we decided to validate our procedure by neglecting the effects due to the accretion disc transition to a slim-disc. In practice: i) we employed 
\begin{equation}
         \log{\frac{R_{\rm BLR}}{\rm 1 lt-day}} = 1.555\pm 0.31 + 0.542^{+0.035}_{-0.033}\log{\frac{L_{\rm 5100\mathring{A}}}{10^{44}\rm\, erg\, s^{-1}}}
     \end{equation}
in Eq.~\eqref{eq:fwhmHb}, as done in \citet{rv15}, ii) 
we inferred the broad-line luminosities from the scaling relations in \citet[][see also Eq.s~\ref{eq:LHb} and \ref{eq:LHa}]{bentz13}, using our tabulated value for $L_{\rm 5100\mathring{A}}$, and iii) we assumed a constant $f_{\rm virial}=1.075$ as in \citet{rv15}. With these assumptions, we found our best parameters to be in line with those in the published works, as shown in Fig.~\ref{fig:MBH}. The inset shows the remarkable agreement of our procedure with the data by \citet{rv15}. The only mild discrepancy is in the data by \citet{greene24}, where the estimates show a somewhat larger scatter around the 1:1 relation. The systematic small shift of the \citet[][above]{harikane23} and \citet[below]{maiolino23} is likely related to the information provided in the respective papers.  \citet{maiolino23} report the $H\alpha$ flux, which is then converted into luminosity assuming the cosmology and redshift reported in the discovery paper, while \citet{harikane23} give directly the broad line luminosity. We will refer to the MBH masses obtained with this procedure as `validation' in the following.

In general, we find that the spin is very poorly constrained by our MCMC, due to the limited amount of observational data we have and the moderate dependence on its value, whereas the MBH mass and $L/L_{\rm Edd}$ are typically well determined. As an example of the robustness of our procedure, we report in Fig.~\ref{fig:corner} the corner plot obtained for J1030+0524 from \citet{yue24}, one of the most massive sources in the sample, which is also one of the few validation cases in which the posterior distribution of the MBH spin exhibits a peak rather than being almost flat. The blue lines in the corner plot correspond to the estimates from the literature, which agree well with our estimate. 

\subsection{Full model}
\begin{figure*}
    \centering
    \includegraphics[width=0.98\columnwidth]{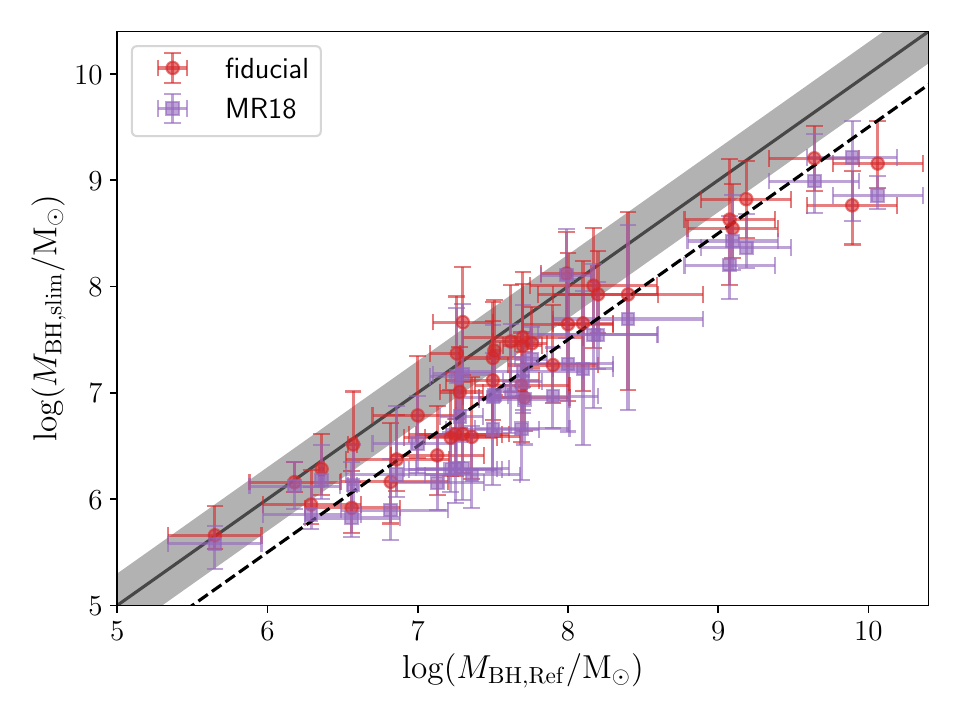}
    \includegraphics[width=0.98\columnwidth]{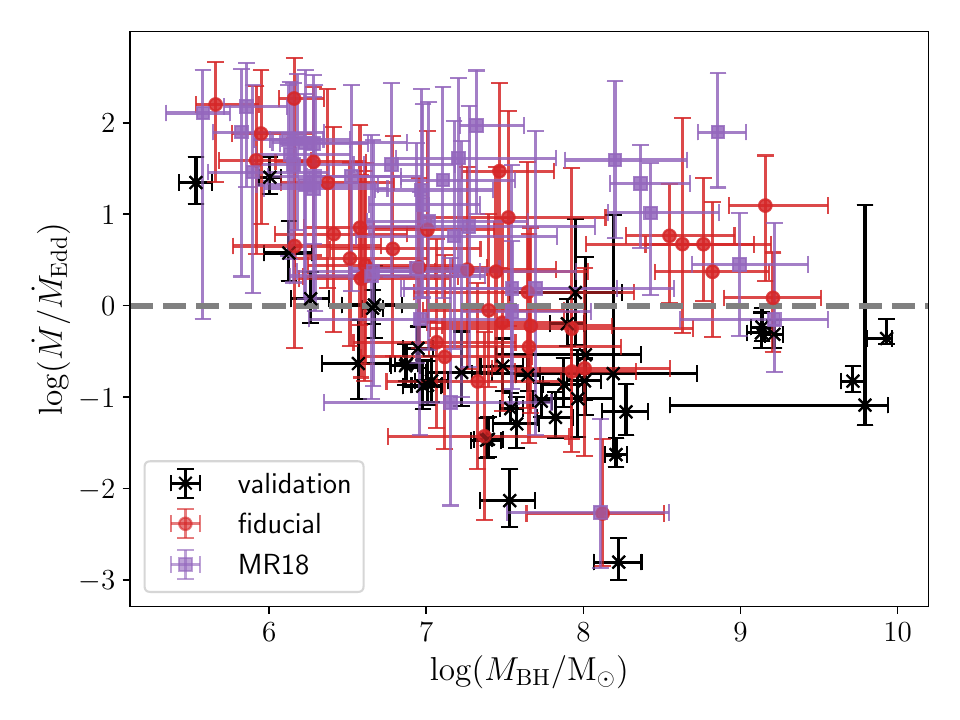}
    \caption{Left panel: same as Fig.~\ref{fig:MBH}, but for our full model, with the estimates of the entire sample shown as red dots (fiducial) and purple squares (MR18). The black dashed line is to guide the eye and corresponds to a 0.5 dex offset relative to the 1:1 relation. Right panel: Eddington ratio distribution for our fiducial model (red dots), the MR18 case (purple squares), and the validation run (black crosses) as a function of the estimated MBH mass. The thick grey dashed line corresponds to the Eddington limit.}
    \label{fig:MBHfull}
\end{figure*}

\begin{figure*}
    \centering
    \includegraphics[width=0.48\textwidth]{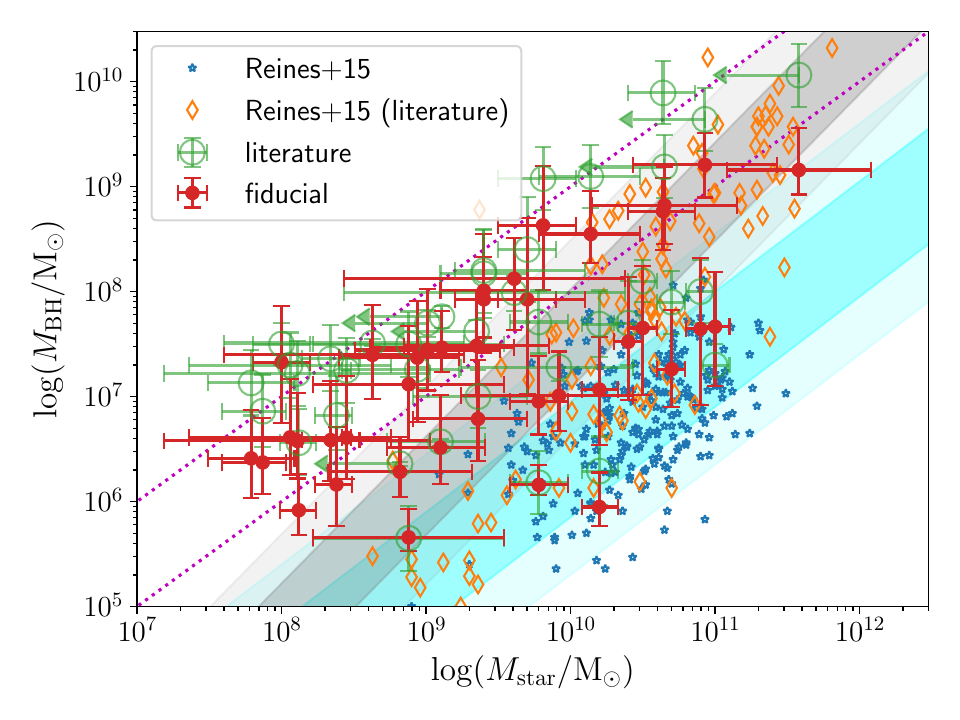}
    \includegraphics[width=0.48\textwidth]{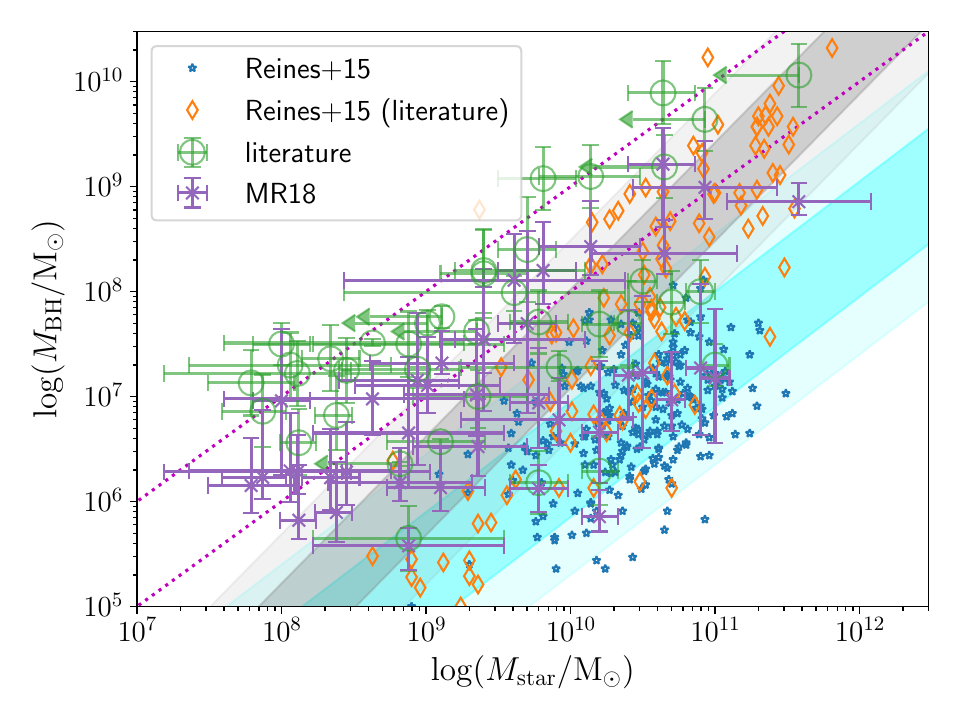}
    \caption{MBH mass--stellar mass relation for the source in our sample. We show the local AGN from \citet{rv15} as blue stars and orange diamonds, with the underlying shaded area correspond to the 1-$\sigma$ and 2-$\sigma$ uncertainties around the best fits to the local samples (grey and cyan for inactive and active galaxies respectively). The original data from the literature is shown as green circles, whereas our new estimates are reported as red dots (left panel) and purple crosses (right panel) for the two virial factors considered. For completeness, we also show as magenta dotted lines constant mass ratios of 0.01 and 0.1.}
    \label{fig:correlation}
\end{figure*}

In the left panel of Fig.~\ref{fig:MBHfull}, we show the same plot of Fig.~\ref{fig:MBH}, but for the slim-disc model. We clearly observe that the fiducial case is close to the 1:1 relation, but typically offset of about 0.5 dex towards lower values compared to those reported in the literature, with correspondingly higher accretion rates, often super-Eddington. The MR18 case, because of the additional dependence of the virial factor on the broad-line $FWHM$, results in even lower MBH masses. The $\dot{M}/\dot{M}_{\rm Edd}$ ratio is shown in the right panel, where $\dot{M}_{\rm Edd}\equiv 10L_{\rm Edd}/c^2$, assuming the fiducial and the MR18 cases of our slim-disc model (red dots and purple squares respectively) and the validation run (black crosses). Despite the differences in the two slim-disc models, we find that the distribution of Eddington ratios is quite similar, with the least massive MBHs more often preferring higher accretion rates. The MR18 case, consistently with the lower MBH masses reported in the left panel, almost always prefer super-Eddington rates, with values between 10 and a 100 times Eddington. Interestingly, the most massive MBHs from \citet{yue24} also prefer super-Eddington accretion rates, well above the Eddington limit, which might hint at an ineffective self-regulation of their growth via feedback processes. Another interesting aspect is that, even in the validation run, some MBHs seem to lie above the Eddington limit, especially those with the lowest masses, suggesting that the estimate of their properties according to the local correlations might be biased towards higher MBH masses. Finally, note that discriminating such high accretion rates from more typical cases is not easy, as the luminosity of these objects would never exceed, even in the most extreme cases, a few times the Eddington luminosity (5-10).

As already discussed above, also with slim-disc model, the MBH and the Eddington ratio in our analysis are typically constrained within one order of magnitude, whereas the MBH spin is almost always uniformly distributed, which suggests that our model can accommodate  the observed data almost independently of the spin. At super-Eddington rates this is expected, as the effective radiative efficiency does not depend on the spin \citep{madau14}. Below the Eddington limit, instead, this suggests that the information available is not sufficient to actually disentangle the spin from the other two parameters.

Finally, we can assess how the correlations between MBHs and their hosts would change considering the results of our full model. The results are shown in Fig.~\ref{fig:correlation}, for the fiducial case (left panel) and the MR18 one (right panel).  We can observe that our estimates are closer to the local relations, and this effect is more relevant for the MR18 case. Interestingly, this decrease does not completely realign the MBHs with the local correlations, but suggests that the current estimates, especially for the lowest mass MBHs observed, could have a much larger uncertainties than reported in the literature, and their overmassiveness relative to the host should be considered in the light of what we found in this work, besides observational biases.

Even though not reported here, as an estimate of the stellar mass is not available, we performed our analysis also on the sources by \citet{matthee24}, finding similar variations in the MBH mass to those just discussed. In order to check whether the inclusion of a slim disc emission produced a rigid shift of the MBH masses also for the local sources, we reanalysed the \citet{rv15} sample using our fiducial model. We found that, on average, a decrease in the inferred MBH mass was also present in the local AGN sample, but with variations not larger than 0.2~dex, about a factor of 3 smaller than the intrinsic uncertainty by \citet{rv15}, and typically much smaller than the 0.5~dex found in the high-redshift sample.

\subsection{AGN spectra}
\begin{figure*}
    \centering
    \includegraphics[width=0.98\textwidth]{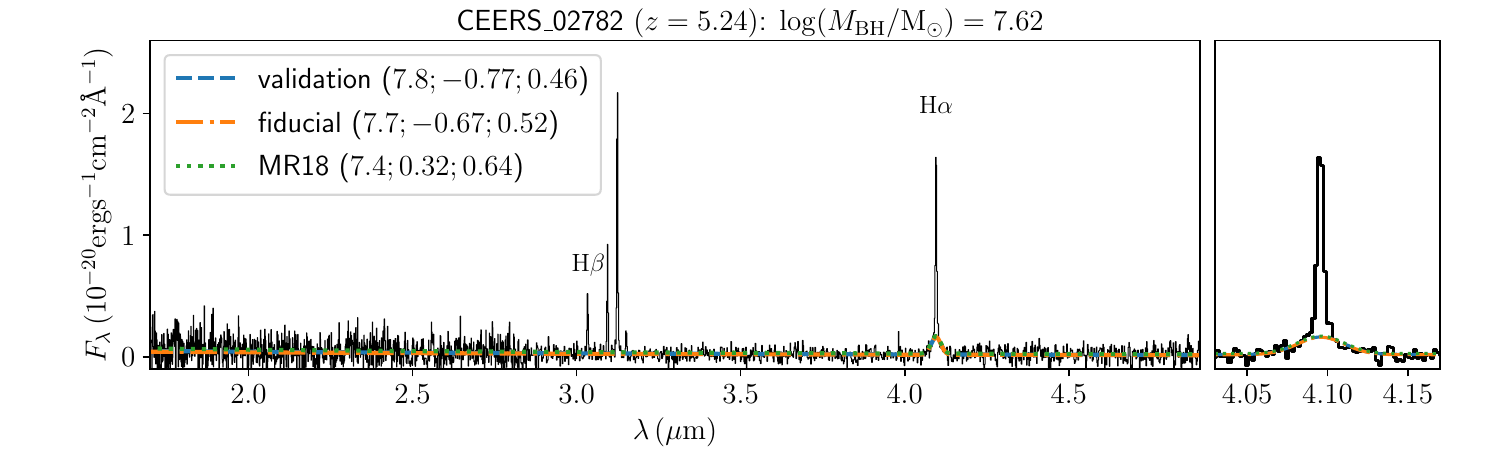}
    \includegraphics[width=0.98\textwidth]{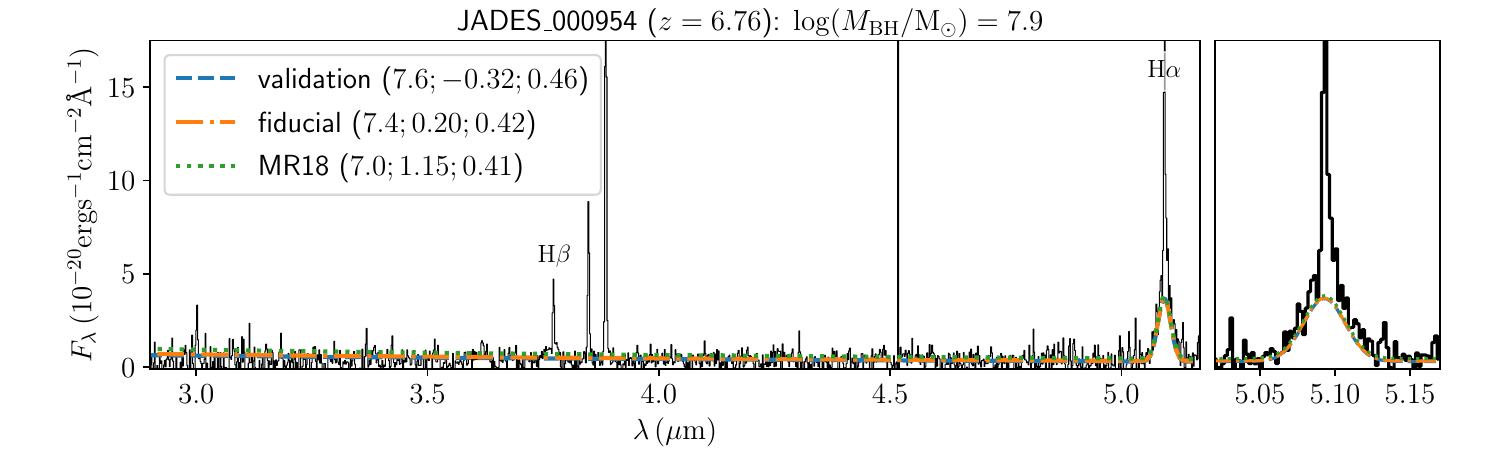}
    \includegraphics[width=0.98\textwidth]{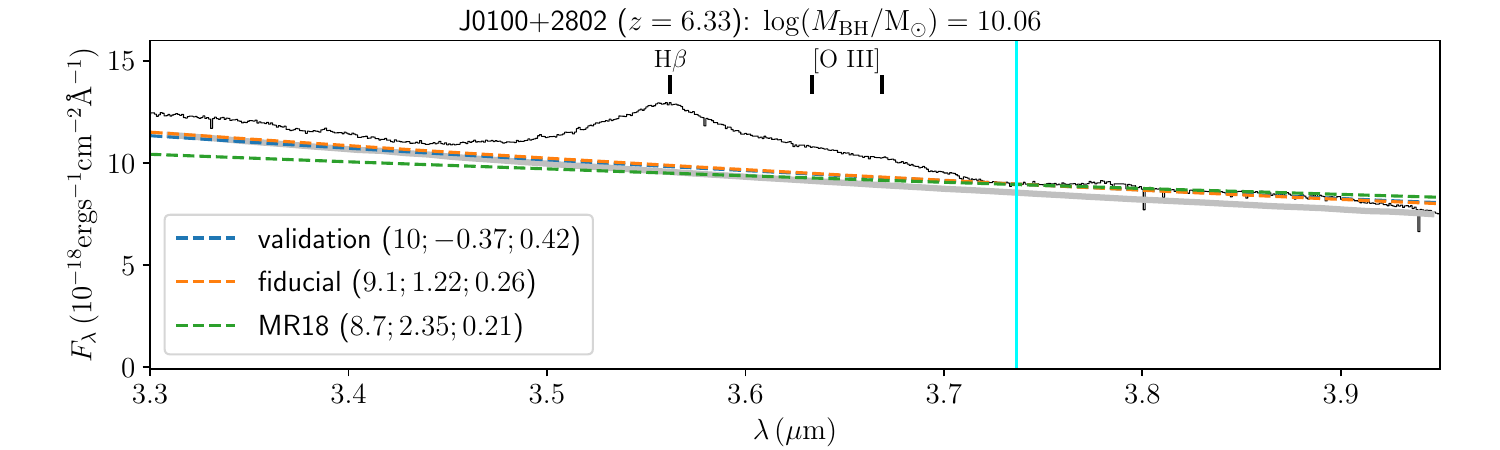}
    \includegraphics[width=0.98\textwidth]{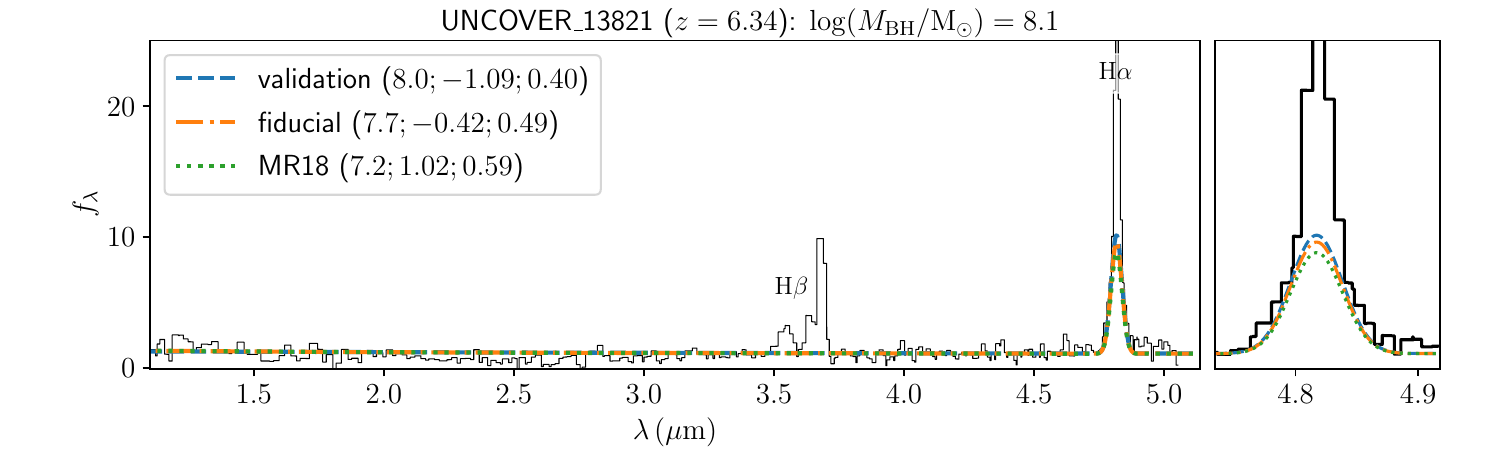}
    \caption{Reconstructed spectra for 4 selected sources in our sample: CEERS\_02782, JADES\_000954, J0100+2802, and UNCOVER\_13821. The observed spectra (obtained from the public data release of the different programs, but for the UNCOVER source, which has been extracted from the published paper) are shown as black solid lines (with the right panel showing a zoom on the H$\alpha$ line), the blue dashed, orange dash-dotted, and green dotted lines refer to our validation, fiducial, and MR18 models respectively. The cyan vertical line in J0100+2802 corresponds to $\lambda=5100\mathring{A}$ redshifted to the observer frame, which we used to constrain the models. The grey line corresponds to the power-law continuum component from the fit by \citet{yue24}. All but the UNCOVER source report absolute fluxes, whereas in the UNCOVER case the flux is normalised to the luminosity at 2500\AA, as done in \citet{greene24}. The numbers reported in the legend correspond to the parameters employed for each model $\log (M_{\rm MBH}/\rm M_\odot)$, $\log (L/L_{\rm Edd})$, and $a_{\rm BH}$, whereas the mass estimates above each panel correspond to those in the corresponding observational papers.} 
    \label{fig:spectra}
\end{figure*}

As a final check of our procedure, we built synthetic MBH emission spectra for the analysed sources employing all of the three models considered in this work. The best parameters to build the spectra are defined as the average among the 10 evaluations of our MCMC with the maximum likelihood. For each model, we extracted the continuum spectrum from our tables and added on top the emission of the broad line (but for the sources in \citealt{yue24}, where we employed the luminosity at 5100\AA). In order to consistently compare with observed spectra, we also accounted for dust extinction following the attenuation law by \citet{calzetti2000}, assuming $R_{\rm V}=4.05$ for the source by \citet{harikane23}, and the Small Magellanic Cloud value $R_{\rm V}=2.74$ for the sources by \citet{maiolino23} and \citet{greene24}, to be consistent with the assumptions in the different studies. For the sources observed by the EIGER program \citep{yue24} we did not include any attenuation. The results are reported in Fig.~\ref{fig:spectra} for 4 selected sources: CEERS\_02782 \citep{harikane23}, JADES\_000954 \citep{maiolino23}, J0100+2802 \citep{yue24}, and UNCOVER\_13821 \citep{greene24}. We clearly see that our models can always recover the spectral properties of the sources, both the continuum region and the broad H$\alpha$ line intensity and width, independently of the assumptions. The only peculiar case is J0100+2802, where the complexity of the broad H$\beta$ line profile, not symmetric and with potential hints of offset components, together with the missing modelling of the Iron emission in our model, does not allow us to recover the exact spectrum. Nonetheless, we find that our model very well reproduces the power-law continuum, but for a mildly higher normalisation, simply due to the use in our MCMC of the total continuum luminosity reported in \citet{yue24} instead of the  contribution of the power-law component only.\footnote{As a check, we re-ran our MCMC on J0100+2802 with a 5\% lower luminosity at 5100\AA\ (consistent with the expected power-law contribution), and found that with almost identical MBH mass estimates the agreement with the power-law fit was remarkable, as expected.} This confirms i) the robustness of our procedure, and ii) that the dependence of the BLR emission on the accretion disc structure and the Eddington ratio is somewhat degenerate, resulting in potentially significant differences in the MBH mass estimate if not properly taken into account.

\section{Discussion and conclusions}
\label{sec:discussion}
In this work, we have built a semi-empirical model of the BLR emission of MBHs in different accretion regimes. By combining theoretical models of the emission of thin and slim accretion discs 
\citep{kubota19} with observed scaling relations at low-redshift which naturally account for different accretion regimes, we have built a versatile model that can be applied to high-redshift sources as those recently observed by JWST. 

We have incorporated our model in a MCMC tool that we used to re-analyse some recent candidate MBHs from JWST observations. Our results showed that, in many cases, a super-Eddington accreting MBH is preferred with respect to the standard SS accretion disc, which translates in MBH masses of up to an order of magnitude lower. This is in contrast with local sources as those by \citet{rv15}, where more than 95 per cent of the AGN are sub-Eddington and our fiducial model almost perfectly recovers the masses reported in the literature. We also note that the missing detection in X-rays of many of these sources might be compatible with a slim accretion disc, but we leave this aspect to future investigations.

Despite the extreme relevance of potentially detecting and identifying highly super-Eddington sources, the sustainability of this accretion phase over long time-scales is unclear \citep[see, e.g.][]{regan19,massonneau23,lupi24}. In particular, there is a potential degeneracy between the MBH mass and the Eddington ratio, and we cannot completely exclude a biased preference for super-Eddington accretion in low-mass systems. In fact, because of the radiation trapping in the innermost regions of the accretion disc, which suppresses the increase in ionising and bolometric luminosity, a slim disc model has more freedom to match the combination of $FWHM$ and luminosities of some of these sources compared to a standard SS disc, without being for this reason more physically plausible. Moreover, any difference in the structure of the BLR (different geometry of the clouds, different density, etc.), as well as different inclinations, might in principle produce similar effects without requiring a highly super-Eddington accretion rate. All these uncertainties enter the virial factor, whose definition can produce variations in the MBH mass estimate up to one order of magnitude, as we have shown here, especially in high-redshift systems for which only a limited amount of information is available.

As for our model, \citet{king24} pointed out that high-redshift MBH mass estimates could be biased toward too high values. Differently from \citet{king24}, in our analysis we did not consider any radiation beaming nor the possibility that the BLR might be mainly dominated by unvirialized outflows. Considering the more likely super-Eddington nature of many observed sources, and the fact that in these conditions radiation beaming as well as nuclear ouflows becomes more significant, we expect the uncertainties in the mass estimate to become even larger. Unfortunately, the limited data available does not allow us to confirm whether a bias in the mass is real or not, and whether such a bias might realign MBHs with the local correlation. However, it provides some insights on the impact of detailed accretion disc physics on the MBH mass estimates. In the future, we will incorporate additional information from the observed spectra that will help us to better constrain the actual mass through our physically motivated model.

\begin{acknowledgements}
AL and AT acknowledge support from PRIN MUR "2022935STW". AL thanks the organisers of the "Massive black holes in the first billion years" conference Micheal Tremmel and John Regan, as well as Ricarda Beckmann, Amy Reines, and Alberto Sesana for useful discussions that inspired this work.   
\end{acknowledgements}

%-------------------------------------------------------------------
\bibliographystyle{aa}
\bibliography{biblo}

\end{document}